# SELF-MAGNETIZED ELECTRONIC FILAMENTS

C.P. Kouropoulos

kouros@bluewin.ch

## *Abstract*

In the one-dimensional world of a flux line beyond $B_{c\_e}$ = *4.41GT* paired evanescent low momentum 0-Landau electron states condense into magnetized bosons of vanishing momentum. I use the results of Rojas et al. within a potential and show that magnetic intensities and electron densities sufficient to initiate condensation are achieved in the shockwaves of cathode hot spots; from unipolar pulses on the surface of a discontinuous micro conductor or cathode tip as a discharge suddenly self-interrupts. Above $B_{c\_e}$, there are charged electronic vector states that are superconducting and ferromagnetic. Assuming the magnetization of the vacuum within the flux quantized state to be free thanks to the Euler-Heisenberg renormalization extends the range of allowed stable states. The thinnest filaments have the Compton radius for the effective boson mass. When perturbed, they evolve towards a decaying state with negative mass and vanishing density at $B_c$ or further condense into composites of higher mass, charge and critical field. Open, almost neutral filament systems with one flux quantum and longer than four microns would be metastable in the Earth's magnetic field and terminated by magnetic monopoles.

## Introduction

Although the effects of magnetized electronic condensates have been observed for decades in the form of charge clusters, monopole pairs and sources of low energy nuclear reactions, they are a neglected aspect of condensed matter physics, mostly because they imply extreme conditions of magnetization and charge density that were expected to be relevant only to astrophysics. In fact, these may commonly occur in various arc and spark phenomena. The resulting condensates turn out to be conducive to efficient micro-fusion and nuclear transformations. The energy stored in them partly originates from the forced charge separation that occurs after their formation, when part of the screening condensate of confined nuclei is dispersed against the binding Coulomb potential by its nuclear reactions. First, the results of Rojas et al will be reviewed; then applied to the description and dynamics of such condensed electronic states. A consequence of the anomalous magnetic moment is the possible existence of states with a negative mechanical mass. These would represent only a tiny portion in the spectrum of such states, but would have observable consequences for the overall dynamics, as well as a distinct radiative signature in the form of pairs of $E \leq \gamma_A M \sim 1.18 keV$ photons.

## Magnetic bosons

A vanishing momentum electron gas becomes one-dimensional and can be bosonized when the magnetic binding of the 0-Landau state exceeds that of the ground Coulomb state *(n=1)*

$$mc^2\left(1-\sqrt{1-e\hbar B/m^2c^3}\right) \cong \frac{e\hbar B}{2mc} > \frac{(Ze^2)^2 m}{2\hbar^2 n^2} \quad . \quad (0)$$

*i.e. B > 235000 $Z^2$[T]*, with Z that of the heaviest nuclei in the system and $Z=\sqrt{2}$ for (virtual) positronium in vacuum. In the one-dimensional world of a magnetic flux line beyond the critical field $B_{c\_e}$ = *4.41GT*, such electrons acquire a complex energy and bosonize spontaneously: evanescent electron pairs condense into charged vector bosons whose charge, spin, mass and magnetic moment are assumed to be the sum of their constituents'. This fails to occur in higher energy scattering, because of the *$p^2$* term under the square root and simultaneous decrease of the ordinary magnetic moment with increasing momentum.

## *The dynamics of condensation*

The charged vector bosons in $n = 0, 1, 2...$ Landau orbits about the z-axis magnetic field have energy eigenvalue

$$\varepsilon_{n^{\pm}} = \sigma c \sqrt{(M_{n^{\pm}} c)^2 + p_z^2} + U \tag{1}$$

$$\sigma = \begin{cases} +1, & M_{n^{\pm}} \geq 0 \\ -1, & M_{0^-} < 0 \end{cases}$$

In addition to the electrostatic energy, $U$ may include the magnetic energy per pair. While the mass per boson for the overall condensate is $\varepsilon_{n^{\pm}}/c^2$, the boson effective mass within the condensate and transverse kinetic term (relative to the magnetic field) is

$$M_{n^{\pm}} = M\left\{\sqrt{1 + (2n \pm 1)x^2} \pm \gamma_A x^2\right\} \tag{2}$$

For simplicity, it was and will be assumed hereafter that the boson has the additive properties of its constituent fermions so that

$$M = 2m_e, \quad Q = 2e, \quad \gamma_A = \frac{1}{2}(g - 2) = \frac{\alpha}{2\pi} + ...$$

and

$$x^2 = \frac{2e\hbar B}{M^2 c^3} = B/B_c. \qquad \text{b)}$$

The 0-Landau state, which alone is allowed for a neutral boson, has

$$M_{0^-} = M\left(\sqrt{1 - x^2} - \gamma_A x^2\right), \tag{3}$$

with the energy

$$\varepsilon_{0^-} = \sigma c \sqrt{(M_{0^-} c)^2 + p_z^2} + U. \tag{4}$$

Above $B_c \sim 8.81$ GT, both $M_{o^-}$ and the Hamiltonian become complex and the state decays. The thermodynamic potential $\Omega = -\beta^{-1} Log\, Z$, with $\beta = 1/k_B T$, involves the chemical potential $\mu$ that defines the overall state.

$$\Omega = \sum_n \Omega_n \tag{5}$$

$$= \frac{1}{2\pi^2 \hbar \lambda_B^2 \beta} \sum_n b_n \int_{-\infty}^{\infty} dp_z \, \ln\left[\left(1 - e^{-(\varepsilon_n^{\pm} - \mu)\beta}\right)\left(1 - e^{-(\varepsilon_n^{\pm} + \mu)\beta}\right)\right]$$

with quantum magnetic radius $\lambda_B = \sqrt{\hbar c / eB}$. The one-loop expression for the density of the charged vector ground state that contributes the most or neutral boson state is

$$N_0 = -\partial_\mu \Omega_0 \tag{6}$$

$$= \frac{1}{\pi^2 \hbar \lambda_B^2} \int_0^{\infty} dp_z \left( \frac{1}{e^{(\varepsilon_{0^{\pm}} - \mu)\beta} - 1} - \frac{1}{e^{(\varepsilon_{0^{\pm}} + \mu)\beta} - 1} \right).$$

Since only the lowest power in the exponential is significant, this approximates as

$$N_0 \cong \frac{1}{\pi^2 \hbar \lambda_B^2 \beta} \int_0^{\infty} dp_z \left( \frac{1}{\varepsilon_{0^{\pm}} - \mu} - \frac{1}{\varepsilon_{0^{\pm}} + \mu} \right) \tag{7}$$

$$= \frac{1}{\pi^2 \hbar \lambda_B^2 \beta} \int_0^{\infty} dp_z \, \frac{2\mu}{\varepsilon_{0^{\pm}}^2 - \mu^2}$$

Since $N_o = \Sigma_{\pm} N_{o^{\pm}} \simeq N_{o^-}$,

$$N_o \cong \frac{2}{\pi^2 c\hbar \lambdabar_B^2 \beta} \left\{ \frac{\mu - U}{\sqrt{M_{o^-}^2 c^4 - (\mu - U)^2}} \arctan \frac{M_{o^-} c^2 + (\mu - U)}{\sqrt{M_{o^-}^2 c^4 - (\mu - U)^2}} \right.$$

$$\left. + \frac{\mu + U}{\sqrt{M_{o^-}^2 c^4 - (\mu + U)^2}} \arctan \frac{M_{o^-} c^2 - (\mu + U)}{\sqrt{M_{o^-}^2 c^4 - (\mu + U)^2}} \right\} \quad (8)$$

Let $\nu = \mu - U$, $\nu = \mu' + M_o c^2$, $\mu' \to 0^-$.

$$N_o \cong \frac{2}{\pi^2 c\hbar \lambdabar_B^2 \beta} \left\{ \sqrt{\frac{M_{o^-} c^2}{-2\mu'}} \arctan \sqrt{\frac{2M_{o^-} c^2}{-\mu'}} \right.$$

$$\left. + \frac{M_{o^-} c^2 + 2U}{\sqrt{-UM_{o^-} c^2 - U^2}} \arctan \frac{-U}{\sqrt{-UM_{o^-} c^2 - U^2}} \right\} \quad (9)$$

At equilibrium, $N_o$ is arbitrarily large, as is the first term in (9), while the second is real and bounded when $0 \leq -U < M_o c^2$, with $-U$ not too close to $M_o c^2$. Else, the state acquires a complex energy density $N_o M_o c^2$ and decays. Otherwise

$$N_o \cong \frac{1}{\pi \hbar \lambdabar_B^2 \beta} \sqrt{\frac{M_{o^-}}{-2\mu'}} \,, \quad (10)$$

then,

$$\mu' \cong -\frac{M_{o^-}}{2} \left( \frac{1}{\pi N \hbar \lambdabar_B^2 \beta} \right)^2 . \quad (11)$$

Using (6) and (8)

$$\Omega \cong \frac{1}{\pi c \hbar \lambda_B^2 \beta} \sqrt{M_{o^-}^2 c^4 - \mu^2} \qquad (12)$$

Also, from the approximate relation at equilibrium

$$N_o = \frac{1}{\pi c \hbar \lambda_B^2 \beta} \frac{\mu}{\sqrt{M_{o^-}^2 c^4 - \mu^2}},$$

we have

$$\mu = \frac{M_{o^-} c^2}{\sqrt{1 + \left(\frac{1}{N_o \pi c \hbar \lambda_B^2 \beta}\right)^2}}. \qquad (13)$$

## *Charged states in a potential*

The solution for the condensate is $\mu$ minus $U$ on the left hand of (13) with $U$ vanishing or moderately confining as in (9), i.e. $-U \in (0, M_{o^-} c^2[$. The condensate is most stable at $-U = M_o\text{-}c^2/2$, for which the second line of (9) vanishes. Both bounds of the allowed interval are unstable against small perturbations. As will be seen, composite bosons self-condense and magnetize spontaneously, even below their $B_c$. The limiting factors being the density and the screening, in a long solenoid for bosons of order $k>1$, i.e. $M=2km_e$, $Q=2ke$, underscreening ($U > 0$) induces $k \to k-1$, while overscreening ($-U \geq M_o\text{-}c^2$) promotes $k \to k+1$. In this way, the condensate fits the screening.

## *Negative mass states*

When $\varepsilon_{o^-}$ becomes negative from the Coulomb effect of screening nuclei, the electron condensate acquires a negative mass, but its $M_{o^-} > 0$ bosons still behave in a standard way within it. Just beneath $B_c$, $M_{o^-}$ in (3) as well as $\mu$ in (13) are negative. Let $\nu = \mu - U$, with $\nu < 0$ and $\nu = \mu' + M_o c^2$ as before. We have $0 \leq U < -M_o c^2$. Then (11) and (13) result in the infinitesimal $\mu'$ being positive. So the density of states (10) remains positive.

## The energy of the vacuum

$$\Omega_V = -\frac{1}{4\pi \lambda_B^2} \sum_{n=0}^{\infty} \alpha_n \int_{-\infty}^{\infty} dp_z \, \varepsilon_n$$

Here, $\alpha_n = 2 - \delta_{0n}$. Taking into account the creation of electron-positron pairs in strong fields, Euler and Heisenberg found an expression from which

$$\Omega_V = \frac{eB}{8\pi^2 \lambda_B^2} \int_0^{\infty} e^{-y m^2 c^3 / e\hbar B} \left[ \frac{\coth y}{y} - \frac{1}{y^2} - \frac{1}{3} \right] \frac{dy}{y} \quad . \quad (14)$$

Above $B_{c\_e}$, the Euler-Heisenberg regularization adds a large negative term proportional to $B^2$ that absorbs the positive classical energy $B^2/8\pi c$. That is, the vacuum becomes ferromagnetic for $B \simeq B_{c\_e}$. From that, the *free* magnetization of the vacuum is

$$M_V = -\partial_B \Omega_V \quad , \quad (15)$$

Note the insistent presence of the "magnetic" unit quantum flux radius $\lambda_B = \sqrt{\hbar c / eB} \equiv \hbar/M_o\text{-}c$, the Compton radius when, as above, the anomalous magnetic moment is neglected. The importance of $\lambda_B$ leads me to conjecture that only for the inner vacuum of the flux quantized highly magnetized collective state above $B_{c\_e}$, does the renormalized magnetic energy vanish, resulting in a series of potential wells dense towards $B_c$. That is, unquantized flux states allow no renormalization.

## The vacuum pressure

It appears in the diagonal spatial terms of the vacuum energy-momentum tensor

$$T_{V\mu\nu} = 4 F_\mu^{\ \rho} F_{\nu\rho} \frac{\partial \Omega_V}{\partial F^2} - \delta_{\mu\nu} \Omega_V \quad (16)$$

Its transverse component is negative and equal to

$$P_{V\perp} = -\Omega_V - BM_V \quad (17)$$

$$P_{V\perp} = \frac{eB}{8\pi^2}\left\{\frac{1}{\lambdabar_B^2}\int_\varepsilon^\infty e^{-ym^2c^3/e\hbar B}\left[\frac{\coth y}{y} - \frac{1}{y^2} - \frac{1}{3}\right]\frac{dy}{y}\right.$$
$$\left. + \frac{1}{\lambdabar_{C_e}^2}\int_\varepsilon^\infty e^{-ym^2c^3/e\hbar B}\left[\frac{\coth y}{y} - \frac{1}{y^2} - \frac{1}{3}\right]dy\right\} \quad (18)$$

$$\lambdabar_{C_e} = \frac{\hbar}{m_e c}$$

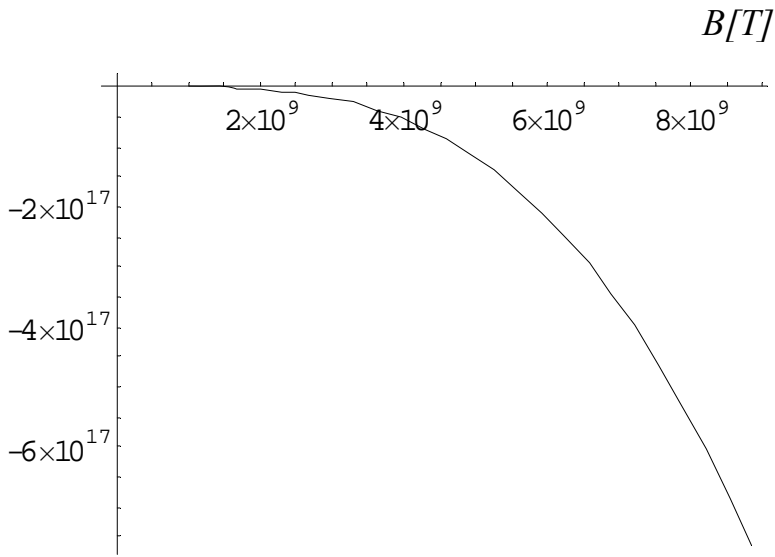

$P_{V\perp}\ [N/m^2]$          Fig 1a

The squared inverse radii are for a unit of quantum magnetic flux, in Gauss units, and the Compton radius respectively. The pressure along the magnetic field is

$$P_{V|} = -\Omega_V \quad (19)$$

and involves only the first term in the brackets (18).

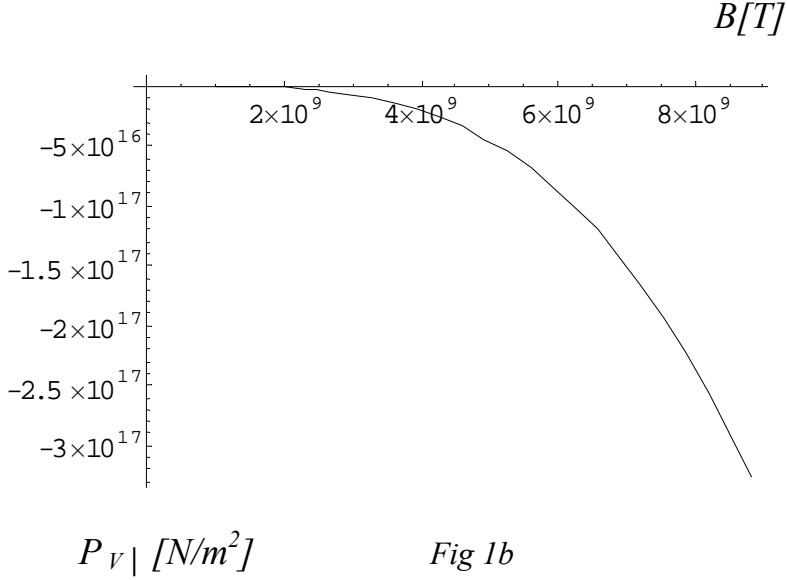

$P_{V\perp}$ [N/m²]   Fig 1b

For *B* = *1 GT*, the negative transverse pressure is *-3 x 10¹⁴ N/m²*, a negative energy density of *– 3 x 10¹⁴ J/m³* that overcomes the positive pressure of the dense Fermi gas in astrophysical objects and induces their collapse. In terms of the negative vacuum energy *per pair* inside a typical self-magnetized condensate, though, even at *2B_c/3=5.88 GT*, that negative vacuum energy represents only *20 meV* per pair, which can be neglected in the energy balance.

## *The magnetization*

In Gauss units, the magnetization inside the condensate is given by

$$M = -\partial_B \Omega$$
$$= (-\partial_\mu \Omega)\partial_B \mu \Big|_{\mu=M_o c^2} \qquad (20)$$
$$= \frac{Ne\hbar}{Mc}\left(\frac{1}{\sqrt{1-B/B_c}} - 2\gamma_A\right) \quad .$$

For $M_{o^-} > 0$ and *B* smaller than $B_c$ and not too close to it

$$M \cong \frac{Ne\hbar}{M_{0^-}c} \quad . \qquad (21)$$

The magnetic moment in (20) does not diverge when $M_{o^-} = 0$, only when $B = B_c$. A consequence is the runaway increase in the magnetization and magnetic moment as $M_{o^-}$ reaches its lowest real value. Also, in Gauss units

$$B = H_{ext} + 4\pi M$$
$$H_{ext} = 0 \qquad (22)$$
$$B = 4\pi M \ .$$

Using (20) and (2b), the last line in (22) gives

$$(x^2 + 2\gamma_A A)\sqrt{1-x^2} = A \qquad (23)$$

$$A = \frac{8\pi N e^2 \hbar^2}{M^3 c^4} \qquad (24)$$

If we omit the anomalous term in (23), we find that, for $x$ to be real, $0 \leq A \leq 2/(3\sqrt{3})$. There are two solutions. One has $x^2 \leq 2/3$ so the density of states $N$ increases with the magnetization and is maximal at $B = 2B_c/3$ and $A = 2/(3\sqrt{3})$. This represents a path to self-condensation. Note the steepness of the magnetization close to $A = 2/(3\sqrt{3})$.

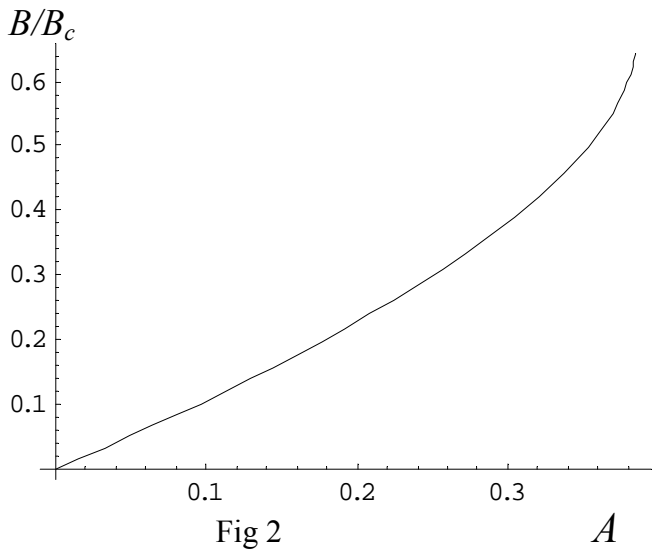

Fig 2

The other solution has $N = 0$ at $B = B_c$ with $x^2$ decreasing from $1-\varepsilon$ to $2/3$ at $A = 2/(3\sqrt{3})$. Since at $B = B_c$ the magnetic moment becomes infinite and the density of such states infinitesimal, the overall magnetization may be lowered by bound Fermi electrons in $n > 0$ Landau orbits, so that the magnetic field does not exceed $B_c$. But the equation of the state beyond $2B_c/3$ is just that for the condensate itself, with no assumption about an additional Fermi fluid. The runaway magnetization and increase in magnetic moment are obviously related to intrinsic changes in the properties of the magnetized bosons the condensate is made of, namely the vanishing of their mechanical energy and effective mass. An interpretation of this solution and of the following curve is that *as the effective mass and equivalent "spin current"(27) of the bosons decrease* (are divided by some factor $\Lambda$), their intrinsic period, *magnetic moment, Compton radius all blow up* (are multiplied by that same factor $\Lambda$), so that they occupy a larger and larger transverse surface, while their longitudinal packing increases somewhat ($d \rightarrow d/(\Lambda-\Lambda^{-1})$) as their magnetization becomes $(1-\Lambda^{-2})B_c$, which decreases the density as $(\Lambda^{-1}-\Lambda^{-3})$. Classically, this is akin to the tendency of the coils in a solenoid carrying a high current to stick together while they expand radially into a thinner and thinner equatorial ring. How to reconcile this blow up with the negative pressure? One answer is that as a larger volume of space is magnetized, a larger volume of classical magnetic energy $B^2/8\pi c$ is absorbed by the quantum regularization of the vacuum energy so that this expansion is accompanied by negative work. There is also the missing electron mass $(M-M_o-)c^2$ from the magnetization. In the vicinity of $B_c = 8.82$ GT, the effective mechanical mass becomes negative and equal to $-\alpha M/2\pi$, which is appropriate to the negative pressure, while its magnetic moment diverges as the state decays. Screening, an additional charged Bose or Fermi fluid as well as the partial renormalization of the magnetic energy beneath $B_{c\_e}$ or owing to a non-quantized flux all affect the mass. During the radial expansion that follows $B > 2B_c/3$, such a state may interact with ordinary electronic states whose reaction to this sudden magnetic impulse can be strongly diamagnetic, from $n > 0$ Landau states, so as to confine the blast if these are external, or damp it if they lie within it. A condensation of additional states inside the blast is also possible. In a white dwarf, the first phase of the magnetization would imply a collapse of the electronic fluid up to $2B_c/3$, and then its runaway radial expansion, while the electron-depleted core undergoes a Coulomb explosion, enhanced by the flight of negative mass electron states. This could be the origin of the highly energetic protons and nuclei often detected in cosmic rays. In neutron stars, the condensate may, as it blows up, initiate the magnetization of the nuclear matter. A competing phenomenon is the further condensation of the vector bosons into composites with higher mass, spin, charge and critical field.

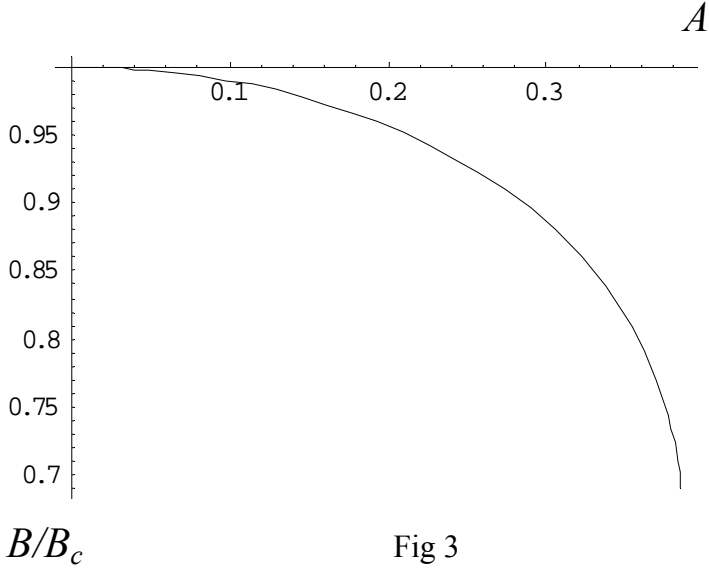

$B/B_c$     Fig 3

The inclusion of the anomalous term in (23) does not change these two solutions noticeably and was taken into account in the preceding graphs.

## *A torus with N' electron pairs*

From (24), the bulk density of the condensate is

$$N = \frac{M^3 c^4}{8\pi e^2 \hbar^2} \frac{x^2 \sqrt{1-x^2}}{1 - 2\gamma_A \sqrt{1-x^2}} \qquad (25)$$

Its lineal and surface counterparts $N_d$ and $N_s$ for an open or closed solenoid (torus) are found by multiplying $N$ by $\pi \lambdabar_c^2$ and $\lambdabar_c/2$ respectively, where $\lambdabar_c = \hbar/M_o\text{-}c$. The filament, a closed flux ring of large radius $R$ with $N'$ bosons, is locally approximated by a long solenoid where

$$B = \frac{4\pi\mu}{c} \nu I,$$

$$\nu = \frac{N'}{2\pi R} \quad .$$

(26)

The current $I$ in each coil (boson), assuming the effective Compton radius for $M_{o^-}$ in (2b), is

$$I = \frac{|\vec{\mu}_{eff}|}{\pi \hat{\lambda}_{eff}^2} \qquad (27)$$

$$= \frac{eMc}{\pi \hbar}\left(\frac{1}{\sqrt{1-x^2}} - 2\gamma_A\right)\left(\sqrt{1-x^2} - \gamma_A x^2\right)^2 \quad .$$

The maximum volume density of condensate pairs is at $N'/R = 6.44 \times 10^{14} [m^{-1}]$, the maximum electrostatic potential per pair on an (unphysical) unscreened torus with $r = \hat{\lambda}_{eff}$ being

$$U_{e\,max} = \frac{2e^2 N'}{C} \quad , \quad C = \frac{\pi R}{\ln(8R/r)} \quad . \qquad (28)$$

See the appendix for various computed values concerning condensates. Without renormalization, only larger $n$ states with a finite number of pairs classically satisfy $0 \leq -U < M_o \cdot c^2$ with a suitable screening charge distribution. The filaments lose magnetic visibility to their larger surroundings when the Dirac condition $eg = n\hbar c/2$ is met. (They become invisible quantum-mechanically, which they already were to a phase-space averaged collection of classical particle or object trajectories and orientations.) In order to minimize the interaction energy,

$$e\int \vec{B}\cdot d\vec{S} = e\oint \vec{A}\cdot d\vec{x} = \frac{nch}{2} \quad . \qquad (29)$$

For $B = x^2 B_c$ and $B_c = M^2 c^3/2e\hbar$,

$$\hat{\lambda}_{B\,n} = \sqrt{\frac{nc\hbar}{eB}} = \frac{\hbar}{xMc}\sqrt{2n} \quad . \qquad (30)$$

For small flux quanta $n$ for which the anomalous magnetic moment can be neglected, this is the Compton radius for $M_{o^-}$. The candidate filament with the lowest field has $A \sim 1/2\sqrt{2}$, $B = B_c/2$, $N'/R = 3.95 \times 10^{14} [m^{-1}]$, $U_{e\,max} = 14 MeV$ and a total energy well above $Mc^2$ if not screened. The match is between *two* Compton radii for $M/\sqrt{2}$ and *two* quanta of flux, which describes a ferro- or antiferromagnetic dual chain. The states with non-integer flux and therefore substantial energy, or awkward geometries with non-integer or higher multiple Compton radii are the most numerous. For small *n*, the ferromagnetic filaments with unit Compton radii for $M/\sqrt{(2n+1)}$ have $x^2 = 2n/(2n+1)$ ; $n = 1, 2, 3...$. They have increasing magnetization and flux number *n*; the nearest such state from $2B_c/3$ being $4B_c/5$, the former is stabilized by the non-quantized magnetic energy in the gap.

Although open filaments may have a renormalized inner flux tube, the magnetic energy of each of their end "charges" must be counted. It is

$$U_B = \frac{1}{2\mu_o} \int_{r \cong \lambda}^{\infty} B^2 dV \cong \frac{1}{\mu_o} 2\pi x^4 B_C^2 \lambda_{B_n}^3 \quad . \quad (31)$$

The $n=1$ and $n=2$ open states are thus terminated by *40.6MeV* and *126MeV* magnetic pseudo-monopoles, respectively. This energy grows with the flux quantum number *n* and is shared among the members of the state so it can be omitted when its population is large and *n* small.

The open filament with the smallest mass $M_{o^-}$ has shrunk into a thin equatorial ring with $n=1$, and $1-x^2 \simeq 1/743602$, large radius $R = \lambda_c$ that increases with $N'$, potential (28), current (27), and maximum magnetic field on the surface with, in this case, the little radius *r* substituted for *R* in (26) by duality. $N'/r$ is $\sim 10^{30} [m^{-1}]$. The magnetic field at the centre is proportional to $N'$, while the overall density is $2.76 \times 10^{36} [m^{-3}]$, about one hundredth that for the $2B_c/3$ state. It has magnetic energy

$$\varepsilon_M = LI^2/2c^2 \quad , \quad L = 4\pi R[\ln(8R/r) - 2] \quad . \quad (32)$$

In the long extreme solenoid or very large toroidal flux ring, the renormalized magnetic energy reaches $215Mc^2$/pair. Such a state, if not stabilized, is typically transitory, as is that with the lowest effective mass $-\alpha M/2\pi$. The latter has a Compton radius of *1.66Å*, but its magnetic moment diverges as $B_c$ is neared: its bosons are decaying fast, for instance by crossing the $B_c$ threshold, into two electrons tunneling out or by further fusion of bosons.

Classically, the evolution towards greater Compton radii as $M_{o^-} \to 0^+$ is akin to that of a coil under magnetic and Coulomb stresses. When $M_{o^-} < 0$, the Compton radius shrinks again because a negative mass reverses the dynamics, according to R. Forward and H. Bondi. Instead of a state evolving all the way to minimal $M_{o^-}$, its radius dilating and then contracting again once $M_{o^-} < 0$, $M_{o^-}$ may directly tunnel from below $\alpha M/2\pi$ to above $-\alpha M/2\pi$ at constant radius and expel $E \leq \alpha Mc^2/2\pi \sim 1.18keV$ photon pairs.

Charged states with the largest Compton radius and lowest positive effective mass are stabilized by a positive Coulomb cloud of similar radius.

## *How are such filaments formed?*

They result from the runaway magnetization to $B_{c\_e}$ that initiates at the density of matter, from 0-Landau electronic states in a closed flux, ring-like microscopic electronic shockwave. A magnetization sufficient to initiate condensation can be reached from very short, unipolar pulses traveling down a wire or transmission line (ideally, nonlinear and pulse-shortening), when the current spike, already squeezed into the surface by the skin effect in the trailing edge of the pulse's voltage (owing to the self-inductance), reaches a discontinuity such as the tip of a truncated conical cathode; often, a small asperity on its end surface. As the current pulse concentrates within the micro-tip and is suddenly reflected when the latter is pinched and explodes, the magnetic field grows locally and a magnetized electronic shockwave results. A similar phenomenon occurs in cathode hot spots during sputtering, even under a modest DC feed current. An arc first establishes itself between the micro-asperity closest to the anode and the latter, focusing within it all the current; it then melts the micro-asperity and its surroundings; under a steady magnetic pinch combined with the electrical attraction of the charge gathering at the tip, a conducting column of molten metal is extruded from the surrounding molten pool, leaving a crater; as the column rises longer and thinner, down to a diameter of 1 micron and as long as a hundred microns, the current density within it reaches *$10^{12}$ A/m$^2$*. At that point, the sausage instability breaks the liquid micro-wire, whose tips at the gaps explode from vaporization; owing to the latter as well as to magnetization, conductivity falls, while the growing induced curled magnetic field, transverse to the current, ensures the latter is interrupted; as a result, a large current pulse is suddenly reflected within each piece of the suddenly pinched broken conductor, so that the induced magnetic field grows even more sharply. Whereas in a bounded conductor at equilibrium, such as a cube of copper, the conduction-band electrons are continuously reflected off the boundaries, their density is that of motionless, delocalized standing waves, so that no charge density pulse is dynamically reflected off the walls and no motional electric field ensues. Here however, a huge motional electric field **$E$** = -$\partial_t$**$A$** appears that slows the internally reflected electrons and rushes the remainder trailing edge electrons against the magnetized perimeter of the cathode's accidental micro-tip. It also attracts a delayed cloud of positive ions to the cathode, in its axis.

In a cathode hot spot, the electric current is typically *$10^{12}$A/m$^2$*, which implies an equilibrium magnetic field of a few Tesla at most.

From

$$\vec{\nabla} \times \vec{H} = \vec{J} + \partial_t \vec{D} \qquad \text{a)}$$

or (33)

$$B_\perp = \frac{\mu}{2\pi R}\{J_| + \varepsilon\,\partial_t E_| + E_|\,\partial_t \varepsilon\} \quad , \qquad \text{b)}$$

a magnetic field also results from the time-variation of the dielectric displacement, *both from the sudden increase and then decrease of the motional electric field*, as a current pulse is reflected at the tip upon vaporization, *and from a fast increase in permittivity as the dielectric medium at the interface is being highly excited, electrically and magnetically, the magnetic field enhancing the medium's dielectric strength, prior to ionization*. It is well known that the dielectric susceptibility of atoms increases with their volume and as the sixth power of their main quantum number *n*. As the atoms at the boiling interface are excited to *n~10*, their susceptibility suddenly increases by a factor of one million, which results in a jump in the overall dielectric constant of several units in a time of a few nanoseconds, or shorter.

The conductivity tensor in a magnetized plasma is

$$\sigma \propto \begin{pmatrix} \dfrac{1}{1+\omega_c^2\tau^2} & \dfrac{-\omega_c\tau}{1+\omega_c^2\tau^2} & 0 \\ \dfrac{\omega_c\tau}{1+\omega_c^2\tau^2} & \dfrac{1}{1+\omega_c^2\tau^2} & 0 \\ 0 & 0 & 1 \end{pmatrix} , \qquad (34)$$

with $\omega_c = eB/m$ the cyclotron pulsation, $\tau = \lambda/v$ the mean time between collisions (electron-ion or ion-ion), $\lambda$ the mean path that is inversely proportional to the pressure, and $v$ the mean velocity. A highly magnetized plasma in which the Hall parameter $\omega\tau >> 1$ for electrons *and* ions behaves as a *very high-strength dielectric perpendicularly to the field*.

Seeing as this tensor is an inverse function of the magnetic field, the dielectric breakdown is temporarily frozen, the discharge avalanche cannot initiate even if the medium is ionized so that it behaves as an enhanced dielectric. Moreover, the magnetic susceptibility of atoms is known to increase in very high magnetic fields. This, along with increasing numbers of electrons in 0-Landau states, contributes to a higher permeability $\mu$ in (33b). The magnetic filament forms as the electrical field varies. Typically, when the discharge is interrupted on a suddenly pinching melted and extruded asperity

$$A_\| \sim \frac{\mu_o}{4\pi} I_o\, LogR \quad, \quad B_\perp \sim -\frac{I_o\, LogR}{8\pi^2 c_{eff}^2 R\, \Delta t^2} \qquad (35)$$

$$R < \tfrac{1}{2}\mu,\ I_o \sim 1A,\ \Delta t < 1ns$$

Round the micro-tip interface, the intensity of the fast magnetic pulse that surrounds the electronic shockwave exceeds *4MTesla*. Rising times shorter than *1ps* have been reported, which would raise the field beyond *4 TTesla*, not counting the effect of the enhanced permeability and permittivity and their variation; under similar circumstances, five billion Tesla are exceeded when $\Delta t < 28ps$. This is enhanced by microwave oscillations, ideally of period $T \sim 4\Delta t$ or smaller, such as, for instance, from the relaxation of strong ferromagnetic domains, a resonance in the breaking micro circuit or some external source. According to fig. 2, the critical field $B_c$ need not even be reached, seeing as such a system self-magnetizes when $B > 2.35\ 10^5 Z^2$[T]. The electrons in the shockwave tunnel into the adjacent flux ring and become densely bound and paired within it in a 0-Landau state condensate. So we have our ring-shaped, one-dimensional, dense and magnetized electronic shockwave round the micro-tip and the results of Rojas et al. can be applied.

As the condensate forms, surrounding ions screen it while bringing in additional electrons in the *0* Landau state. Such screening is most efficient from boson nuclei, whose pressure is lower and reactivity moderate. At that point, the densely packed, screened nuclei, collectively undergo fusion reactions and exchange interactions as well as beta transmutations, as shown by Filipov, Urutskoev et al. Owing to the production of odd or different nuclei as well as excess momentum, the nuclear condensate may explode, scattering its products and most of the ions. *Such nuclear reactions drive the charge separation and leave a more negatively charged vector ring, unstable if U>0, or, in the presence of an axial magnetic field, a coil that soon unwinds.* One effect of this early nuclear screening is to stabilize the initial electronic state.

This event creates metastable superconducting filaments inside the discharge region with Compton-size little radii and *therefore higher local current densities than are allowed in the ordinary solid state comprised of atoms.* The next round of micro-discharges and cathode hot spots thus involves such electronic filaments, and their breaking or exploding above the superconducting critical field. The induced magnetic fields from such femto-electrodes are high enough to condense higher order or even quark composite bosons.

## *A micro-nuclear reactor*

One condition for nuclear fusion is the transverse confinement of the nuclei. It is met within neutral (or screened) filaments in the 0-Landau state of all spinning nuclei, while *n=1* filaments confine the first Landau state of He nuclei and heavier, because the cyclotron radius $\sqrt{(2\hbar/ZeB_n)}$ is smaller than the Compton radius $\hbar/M_o\text{-}c$ of the bosons for *Z≥2* nuclei and *n=1* quantum of flux ($B_1=2B_c/3$). Filament endings or shrunk states confine the nuclei beneath or near their surface electrostatically. Another condition for fusion that is allowed for by the screening in the newly-formed charged states is a sufficient nuclear density. Charged states also induce the fission of heavy nuclei, especially in the singular electric fields of their endings, or zitterbewegung fluctuations, as well as fast beta transmutations owing to the density of polarized electrons and the availability of low lying final electronic states for the stripped nuclei.

## *The Darmstadt process*

High intensity closed magnetic flux lines also form when two heavy atoms or ions collide at energies ranging from keVs to hundreds of keVs, from the sudden variation of the radial electric field round the centre of mass as the nuclei approach or move away from one another. From the shielded potential

$$qU = \frac{Z^2 e^2}{4\pi\varepsilon_o R} e^{-\frac{R}{R_A}} \qquad (36)$$

and

$$H_o = \frac{1}{2}\mu\dot{R}^2 + qU \quad , \qquad (37)$$

where $\mu$ is the reduced mass, one deduces the relative velocity as the nuclei approach

$$\dot{R} = \sqrt{\frac{2H_o - qU}{\mu}} \qquad (38)$$

and the radius of closest approach where $H_o = qU(R_o)$. The radial electric field at a distance $\rho = R_o$ round the centre of mass, considering the two contributing nuclei and the locally halved shielding distance, is then

$$E_\rho = -2\partial_\rho U(R')$$

$$R' = \sqrt{\rho^2 + (R/2)^2} \qquad (39)$$

$$\dot{R}' = \dot{R}R/4R' \quad , \quad \partial_\rho R' = \rho/R'$$

From which and (38), the adjacent opposite induction flux rings have the intensity

$$B_\rho = \frac{\partial_t E_\rho}{2\pi\rho c^2} \qquad (40)$$

or

$$B = \frac{\mu_o}{16\pi^2} Ze\dot{R}R \left[ \frac{3}{R'^5} + \frac{2}{R_A R'^3} + \frac{2}{R_A R'^4} + \frac{4}{R_A^2 R'^2} \right] e^{-\frac{2R'}{R_A}} \qquad (41)$$

Taking $R_A = 0.25Å$, $Z = 82$, $2\mu = 210$ a.m.u., $H_o = 1KeV$ gives $R_o = 1.16Å$, $B_{max} = 7.12 \; 10^{15} T$ while $H_o = 10KeV$ gives $R_o = .68 \; Å$, $B_{max} = 6.65 \; 10^{18} T$. At an energy of a few dozen keVs, the high field region has a high electronic density as well, which can result in a magnetic condensate. In this case, we have adjacent, dual and opposite field lines. This mechanism produces magnetic fields high enough to condense higher order, mu, or quark composites, rho bosons or deuterons. In the case of a condensing electronic system, such collisions can help nucleate a subcritical field to critical intensity.

Within a high axial magnetic field, the opposing adjacent flux lines join in a wrapping kink at the equator and become open-ended as their intensity decreases further along the axis. Thus, colliding pairs of heavy nuclei in the axis of an electronic filament condensate may kink or disrupt its inner field. In this case, we take $R_A \sim O(\lambda_C)$ in (43) or even $R_A \sim O(r)$ for the extreme ring.

## *Neutral threads*

A metastable neutral state can be a strongly magnetized ($B \sim B_{c\_e}$), self-confining, self regenerating fiber made from confined twirling photons in an electron-positron magnetized plasma with quantized magnetic flux. Rojas et al. showed that such photons acquire mass and a magnetic moment of about 13 times the electron anomalous value. Although some confinement would be ensured by the refractive index of such a system in self-regenerative resonant transition, it is expected to be highly unstable owing to tunneling. The photons sharply decrease the magnetization to mass-energy ratio.

## *Neutral threads of coaxially bound charged states*

Long-lived neutral superconducting magnetized states and quasi monopoles seem to be an experimental fact. Building them from positronium-like entities as well as confined photons poses the problem of their stability. An alternative is to consider bound condensates with different quantum numbers such as a concentric ferrimagnetic configuration built round a higher order electronic core and having a lower order positronic vector sheath.

Assume open, high order electronic filaments *($M=2km_e$, $Q=2ke$, $B_c=2kB_{c\_e}$)* have condensed within an axial magnetic field. The greatest electric potential and gradient being at their tips, more electron-positron pairs will be created there: the electrons condense or are expelled, while positrons wrap themselves round the filament until they sheath it. Initially, this resembles a giant atomic entity. Because the endings of the inner core have the highest electric as well as magnetic fields and gradients, the density of screening positrons will be significantly higher there, condensing into vectors enclosing some of the flux that returns to the other ending via the surrounding space (at their lower critical field), with some flux escaping outwards from the tips. Initially confined to the endings, the positronic condensate spreads to cover the entire filament, so that a ferrimagnetic configuration emerges. It is assumed to be metastable because the positive sheath vectors cannot immediately annihilate with their higher order central counterparts. Nor can the latter form positronium states with external electrons, or the central bosons annihilate with any surrounding "giant atomic" positron clouds. Later on, the binding of nuclei at the endings, as well as an embedding of the slightly negative system into a ferroelectric may further stabilize it.

The negatively charged central core constitutes a trap for nuclei. Because the peripheral positron condensate lowers the overall charge, these keep safer distances from one another, as compared with those entering a newly formed electronic condensate, which partake in fast nuclear reactions and can be ejected, so that a quiet metastable neutral magnetic polymer is possible. Kenneth Shoulders' grey EVs seem to be just such condensate systems involving a positron vector sheathing round a higher order electron core. The many phases and transformations seen in EVs involve partial decays, sheathing, transitions to a lower or higher order, or to a negative mass.

## *The dynamics of negative masses*

Whereas the straight tunneling from the $+\alpha M/2\pi$ to the decaying $-\alpha M/2\pi$ state plus radiation only involves a virtual transitory state, persistent states with $M_o \sim 0^-$ are a conundrum in thermodynamics and causality in approaches that reverse the temporal evolution, and problematic as regards the apparent charge non-conservation in those that transfer such states unto mirror folds with non-interacting charge. The energy is bounded from below at $-\alpha Mc^2/2\pi$, by a limiting state that also happens to have a vanishing density and a limited lifetime.

There are many ways of dealing with negative masses. For instance, one may simply rule them out for states that have exhausted all their positive electromagnetic energy. But this is not necessarily the case. Ultimately, experiment decides which approach is relevant.

After Stueckelberg, Piron defined the generalized Hamiltonian

$$K = \frac{1}{2}\left\{\frac{1}{M_{o^-}}\Pi_\mu\Pi^\mu - M_{o^-}c^2\right\} + W$$

$$\Pi_\mu = p_\mu - \frac{eA_\mu}{c}, \quad \mu = 0, ..., 3 \tag{42}$$

$$\dot{p}_\mu = \partial_{q^\mu}K, \quad \dot{q}^\mu = \partial_{p_\mu}K, \quad d_\tau t = \frac{E-U}{M_{o^-}c^2} = \gamma$$

For the one-dimensional system that concerns us, given the boson effective mass $M_{o^-}$, $dt/d\tau = \gamma \geq 1$ as expected. Negative mechanical masses are acceptable in states whose positive total internal energy evolves towards an extremum that is bounded from below while their charge, electromagnetic fields, temporal and thermodynamic evolution as well as standard causality are preserved. This is consistent with (2), (3) and the other results in this paper. A consequence is the contrarian dynamics of negative masses noted by R. Forward and H. Bondi that is obtained by inverting the sign of $M$ in Newton's $\boldsymbol{a} = \boldsymbol{F}/M$: they accelerate or recoil in the opposite direction in external electromagnetic fields *so that a neutral system of ions or positrons and negative mass electronic states appears to move as though it were entirely positively charged*; it might self-accelerate without the system gaining momentum or kinetic energy; negatively charged entities may cluster together. Internally, instead of a centrifuge acceleration, the spires of a negative mass solenoid feel a centripetal one and tend to separate longitudinally from one another as their magnetic moment diverges. Such states would gravitate toward the Earth and orbit but repel it. The eventual decay of critical $B=B_c$ negative mass states frees its components from bondage and scatters them. At this point, these reverse their masses, velocities and any runaway stops, while charge and momentum are conserved. The renormalized magnetic energy and missing or negative mass is analogous to the classical vacuum that forms inside the core of vortices, which are the visible singular objects, while the missing mass and momentum become delocalized within the surroundings. Upon transition or decay, this energy relocalizes.

## *Extreme rings*

The extreme ($M_o$- ~ $0^-$) magnetic dipole ring ($R=\lambda_c$, $-Q$) is stabilized by a charged speck and some peripheral ions of total charge $Q'<Q$ centered at $R = 0$, so that $U$ is slightly positive, the speck forming with the ring a harmonic oscillator. The stabilizing effect arises from the Coulomb potential well, because the states with increasingly negative masses have smaller radii that resist being reached so long as the positive speck is there, while the $0^+$ state would require a higher central charge for $U$ to be negative. The $0^-$ composites behave roughly as though positively charged in external fields, with a tendency to cluster into their own negative charge.

The $M_o$- $< 0$ closed flux toroid ($r>\hbar/\gamma_A Mc$, $R>>r$, $-Q$) is stabilized against decay by its own negative charge as well as by the nuclei with charge $Q'<Q$ it confines, because of the increasingly smaller little radii of more negative mass states with contrarian dynamics. External fields accelerate them roughly like positively charged massive objects, even though slightly negative. They therefore cluster together spontaneously.

## *Collective effects*

At some point, owing to oscillations of increased amplitude, a stabilizing charged spheroidal or lenticular cloud with internal cohesive forces is drawn far enough from the centre of an $M_o$- ~$0^+$ extreme ring, or $0 < M_o$- $< \gamma_A M$ peripheral toroid (with some screening ions) that such rings may invert their mass, emitting $M_o$-$c^2$ photon pairs, from a negative $U$ having become positive. Owing to the negative mass, they now run away from the positive objects they attract and cluster toward similar negatively charged entities. The oscillating system, besides new motions in external fields, starts to move of its own, chasing itself erratically, oscillating or stopping upon transition or decay. A medium comprised of a large number of similarly oriented mesoscopic oscillators, synchronized by an external electric field and its fluctuations, thus decays into a unidirectional plasma sandstorm with giant charge oscillations and unexpected motions. This is the typical behaviour of Earth lights, Ball lightning and the "EVs" observed by K. Shoulders. When a newly formed, closed flux $2B_c/3$ toroid evolves to a smaller mass as its large radius shrinks, the strong electrical pulse that results induces neutral metallic grains to discharge by ceding electrons to the condensates, thereby gaining a positive charge; it causes a similar collapse in neighbouring rings and orients them so that a system of the large aforementioned oscillators is prepared. Metallic grains may also cede electrons to extreme tori by simple contact or approach, the latter also gathering electrons from the atoms they confine. Upon collapse, metastable extreme tori induce a substantial magnetic cooling. Within the Euler-Heisenberg renormalization scheme, the magnetization of the vacuum above $B_{c\_e}$ is expected to be free. In extreme long solenoid states, this represents up to *215Mc²/pair*. It remains to be seen whether this can actually be tapped as "free magnetic energy".

## *Other considerations*

What is the precise energy balance for a charged metastable state? it is determined by whether the magnetic flux is quantized and strictly confined, or not; the Coulomb screening, and by an additional neutral Bose or Fermi electronic fluid. The latter only decreases the magnetization, because $B > B_{c\_e} = 4.41GT$ forbids *single electron* low momentum 0-Landau states. The longitudinal momentum on an orbit of radius $R$ is

$$p_k = \frac{\hbar k}{2R} \quad . \tag{43}$$

The $k \geq 1$ states in the Fermi fluid are each filled with $k$ bound electrons, with increasing momentum $p_k$, and a magnetic moment decreasing as $\sim 1/\gamma$, the inverse relativistic factor. Because of its high momentum, the Fermi fluid is a source of perturbations and instability to the overall entity, especially under varying external conditions. Its lower magnetization increases the electronic density in the condensate.

For a neutral magnetic filament of length $L$ to remain open, the ambient field at one pole must be stronger than that from the other so that

$$L > \lambdabar_{B_n} \sqrt{\frac{B_n}{B_{ext}}} \tag{44}$$

As a result, open $n=1$, $B_1 = 2B_c/3$ screened filaments longer than about 4 and 30 to 500 microns are stabilized in the field at the Earth's surface and in its immediate vicinity in the solar system, respectively. Similarly to a compass needle, they align their inner field with an external one and strengthen it further along their axis, which allows for EVs-enhanced magnets. The open neutral metastable filament is a pseudo monopole pair, similar to Kiehn's Falaco soliton. The magnetic poles observed by Urutskoiev et al. after strong discharges would then be open screened filaments of several microns in length that move in the magnetic gradients near the ferromagnets that then capture them, rather than free magnetic charges. The limited lifetime of such entities, of about a day within iron, as well as their decay would suggest it, seeing as an *elementary* particle, especially one carrying a magnetic charge, should be stable. It is therefore expected that a material with anomalous magnetization will not leave the typical signature of monopoles when passed through a superconducting loop. Open neutral filaments shorten progressively or decay under the influence of acoustic perturbations that force them to tunnel to $B_c$, so that the longer shielded filaments have the longest lives, which may reach days or weeks in suitable environments. They may also seed the formation of new charged states.

## *Phenomenology*

The monopole endings of open filaments in motion, as well as their negative core can ionize atoms, and thus accelerate their beta decay and retarded neutron emission rates by freeing the required low lying final states.

A charged condensate in a dielectric, gaseous or ionized medium tends to ionize it strongly, as Mesyats et al. have observed in cathode hot spots and during sputtering, or Donets et al. in drift tube, low pressure, reflex electronic discharges. The anode hot spots of wire electric discharge machining are cathode hot spots that form near an anodized layer (of oxide, carbide or nitride depending upon the dielectric), which grows before the discharge establishes itself and self-interrupts, so that the resulting energetic ions are projected against the anode. It is then observed that machined parts from normally nonmagnetic materials, as well as the carbon residue from the oil, become ferromagnetic from the accumulation of magnetic filaments. In fact, a portion of the field of permanent magnets can arise from such an accumulation. Magnetic and electrical circuits or the relative motions of magnets can be optimized to generate copious amounts of such condensates and harness the resulting electromotive or mechanical force, either from HF superficial discharges, or from the combination of materials, texture, Foucault currents, microwave oscillations and condensate channels that act as femto electrode segments in the bulk. Magnetic cooling results when the fields collapse.

Condensation into filaments occurs when z-pinch discharges break up owing to the sausage instability; this creates very localized and dense discontinuities that explain the substantial fusion reaction and neutron emission rates that have been observed when such discharges break up. For instance, M. Rambaut's analysis of such discharges in deuterium concludes to a medium having a fractal dimension of *1.9-1.6*, with $10^3$-$10^4$ electrons in a sphere of atomic radius *.53 Ã*, in agreement with the present paper. Similar phenomena occur in metallic conductors subjected to high current spikes: they fragment from the Laplace pinch pressure as well as from the longitudinally repulsive electrical forces of their positively charged core and negatively charged surface owing to the skin effect, and can then condense magnetized electron pair ring states as the current is reflected off the micro-discontinuities of the freshly broken conductor. Similar occurrences were observed by Mesyats et al. and the

Correas, during the sputtering of metals in vacuum from the small droplet instabilities in the hot spots that appear on a solid plane cathode (thermal energy being released from the nuclear reactions in the newly created EVs), and by Donets et al. in electronic pulsed reflex discharges in low pressure drift tubes, whose pulses, first squeezed by the mirror charges in the drift tube, later form shockwaves and "electron filaments" once reflected at the ends and the nonlinear discharge phenomena having been initiated.

The centre near the axis of the ring at the reflecting tip is electron-depleted, which helps screen the potential, reduces the critical threshold and the density of the accompanying Fermi states.

As the vector condensate forms, the original magnetic field is amplified, and soon, electrons from the conduction band shockwave or from atoms magnetized beyond the critical field of *4.41GT* for single electrons, join it or bind to it. The stripped nuclei, especially when they are bosons, form rapidly a metastable, dense, highly magnetized, electronically screened state in which nuclear reactions and beta transmutations occur, the charged electronic condensate, surrounded by its accompanying Fermi fluid, being a dense source of polarized electrons interacting with highly polarized nuclear targets. The stripped atoms are then expelled as the multiply ionized entities observed by Donets et al.

The emission of energetic nuclei (16MeV protons) from the cathode hot spots of Tesla coil terminals had already been documented by Gustave Lebon at the end of the nineteenth century, and is a typical (though neglected) component of the "ionic wind" bursts that emanate from negatively polarized tips or surfaces. This positive ionic countercurrent heats the cathode on the one hand, hence the aptly named "hot" spots and, on the other, contributes to the observed negative resistance of the discharge by an actual electromotive force.

In the presence of an axial magnetic field, even a relatively modest one, a newly-formed singularity, rather than close into a small ring, may, owing to the flux compression in the current sheet converging toward the pinched singularity (from the conservation of angular momentum in peripheral electrons having high axial Landau states) and to the further amplification of the induced magnetic field by the emerging condensate, form a solenoid that soon uncoils into an open filament under the influence of both its charge and the outer axial field that stabilizes it. It is therefore more active in its initial stages with respect to induced fission of heavy nuclei, pair creation and ionization at its endings and eventually decays into a long monopole-terminated quasi neutral filament, once its charge is shielded.

Interestingly, the observation of monopoles by Ehrenhaft in the 1930's and by Mikhailov since the 1980's involved *particles created by the sputtering of ferromagnetic materials in electric arcs*, while those recently observed by Urutskoev et al. were produced in exploding discharges (titanium sheets in water) and only later bound themselves to ferromagnetic materials. In the observations of Mikhailov an open magnetic filament, oriented in a homogeneous magnetic field, terminated by magnetic poles, one of which is bound to a ferromagnetic particle, will see the magnetization of that pole enhanced. The strength of the apparent magnetic charges when observing a magnetic dipole will depend on its length, its magnetic flux, on the gradient of the external field and whether some flux is enhanced or returned by a nearby ferromagnetic. Urutskoev et al. observed caterpillar-like dual traces in films that could be ions ejected from the poles terminating a large $n$ screened filament. The fact that the anomalous magnetization created by such poles in iron disappears after a few dozens of hours supports the view of a non elementary particle, whose magnetic charge would have otherwise been conserved. It indicates that the filaments linking the poles decay, with lifetime probably proportional to their length. Screened filaments may carry small, longitudinal, reversible magneto-acoustic excitations between their poles, so long as their accompanying magnetization does not exceed $2B_c/3$ by enough to make the state tunnel through the many neighbouring non-integer flux states acting as confining potential barriers. The charged vector condensate filaments are superconducting and ferromagnetic. *Viewed from a distance, the apparent density of charge of a quasi neutral condensate system can be low and similar to that of some moderately shielded electrons.* Because of their low apparent charge, in the quasi absence of a Fermi fluid and in a well-shielded environment, greater acoustic perturbations are allowed before the constituent charged states tunnel to $B_c$. Grigorov et al.'s polaron-embedded, long-lived filaments would be examples. Earlier measurements by G. Egely and G. Vertesy of the anomalous bio-magnetization of various samples showed that metals retained anomalous magnetic properties less than 48 hours, a span confirmed by Urutskoev et al., while wood remained magnetic for about 40 days, in line with Grigorov's experiments. In these, tiny magnetic filaments were first produced in a polymer precursor by high voltage electrical discharges. The viscous dielectric was then electrically polarized by a higher intensity, lower voltage electrical current for several days so as to assemble the small conducting, magnetic filaments into longer threads. Finally, the dielectric was polymerized with its conducting filaments, reinforced by strong currents. Water, because of its high dielectric constant, is particularly well suited to a similar process, whose initiation is expected to occur naturally. When flowing through dielectric rocks or condensing within storms, it can be subjected to strong electric stresses and discharges. The artificial equivalent would be the Meyer cell. Besides, atmospheric water is likely to condense preferably about the magnetic filaments from solar or terrestrial storms, from the Wilson effect of ionizing magnetic pseudo-monopoles in motion, eventually forming magnetic polymers. The final phases of the Grigorov procedure are quite similar to the otherwise incomprehensible "Joe Cell". It is noteworthy that the latter induces both an anomalous conductivity and dielectric breakdown in the liquid, mist and vapour phases of the "charged water", at low and high voltages respectively, the formation of anomalous ferromagnetic residues or "magnecule chains" in initially non-magnetic compounds as well as other anomalies. Because the orbital of a charged scalar round a central ion is generally diamagnetic (only a current-carrying superconducting cyclic molecule ring with positive ions on its periphery could display attractive Laplace forces), because such magnetic polymers appear in a wide variety of materials subjected to or having been in the vicinity of discharges, including dielectrics such as wood, water, hydrocarbons, gases, metals, and because of an anomalously high energy content in subsequent chemical reactions or combustion, the hypothesis of an orbital molecular ferromagnetism must be discarded in favour of nuclei screening a charged magnetic thread. The resulting filaments and pseudo-monopoles, once activated, induce nuclear reactions , the radiolysis of water or fuel as well as enhanced rates of chemical reactions notably in spark-gaps. A similar reasoning applies to the biological ferroelectrics subjected to high local electrical stresses and currents in physiology. This suggests a role for screened filaments and their magnetic polymers in biology that could explain the puzzling biological transmutations observed by L. Kervran in the fifties.

## *Magnetic boson and binding alternatives*

Charged bosons could result from bound states between an electron and a neutrino with parallel spins. Because they are different eigenstates of isospin, Pauli's exclusion principle holds. A.O. Barut found quasi point-like magnetically bound states between an electron and half the spinor components of a Dirac neutrino with a vanishing anomalous magnetic moment, which could explain the parity violation of the weak interactions. Such a vector has $Q=e$ but is expected to have a high mass.

A bound state between a neutrino mixture with rotating electric dipole and an electron at the same Zitterbewegung frequencies but complementary dipole orientations or phase is also expected. The resulting vector has $Q = e$ and $M \sim 2m_e$. In the neutrino theory of light, this amounts to a binding between a highly localized rotating half-photon and an electron.

Other approaches lead to tightly bound electron pairs with parallel magnetic moments at distances below the Compton radius, at which the Pauli exclusion principle no longer holds in strong fields, especially above $B_c$.

One idea that underlies this paper is that the spinning electron behaves as a current in a conducting loop so that two attract one another at close range ($\sim \alpha \lambda_C$). Two similar, isolated commoving point (or spread) charges experience no magnetic attraction and are, in any orbital motion, mutually repulsive magnetically. *Ferromagnetism cannot arise only from the orbital motions of a charge round a positive centre because any such motion is dielectric and diamagnetic*. Therefore, Poincaré's unit charge spinning ring electron of Compton radius, although it has the correct magnetic moment and flux, would be diamagnetic instead of paramagnetic and allow neither ferromagnetism nor tight bound states. It follows from the Dirac equation that the electron spins within half a Compton radius $\lambda_C/2$ at twice the Compton frequency $2m_e c^2/h$. The anomalous magnetic moment results from the orbit being approximated by the edge of a Möbius strip of width $\sim 2\alpha\lambda_C$. The half Compton radius shrinks with the relativistic factor and momentum, unlike the width of the strip, their associated magnetic moments and fluxes becoming comparable at the strong scale. While the poloïdal flux is a relativistic invariant of one quantum, the toroidal flux of the electron is only $(2\alpha)^2$ that value at rest. The gyromagnetic factor of two that induces the correct magnetic moment involves a current twice that of a unit charge: owing to Zitterbewegung, the charge spins in space, accompanied by positive and negative energy oscillations at the frequency $2m_e c^2/h$. In other words, half a magnetized *para*-pair in the 0-Landau state contributes to the current and magnetic moment, without affecting the spin. This doubles the effective current. Because of the counter-rotating positive current, additional balancing negative charge and mutual apparent Lorentz contraction of the counter-rotating sources of retarded potentials, an attractive Laplace magnetic force results, as within and between conductors. This is why collinear electronic spins, magnets and electromagnets attract one another and why the Coulomb repulsion between electrons is overcome at close range.

Acknowledgements: thanks to L. Urutskoev for many interesting discussions concerning his remarkable experiments and to H. Lehn for having drawn my attention to Rambaut's papers and to interesting references and documents concerning Earth lights and the work of Mesyats.

A. Peréz-Martinez, Samina S. Masood, H. Peres Rojas ; *Effective magnetic moment of neutrinos in strong magnetic fields*, Rev. Mex. Fis. 48(6), 501 (Dec 2002)

Qui Hong Hu; *The nature of the electron* , physics/0512265

TIMELINE

1925 Samuel A. Goudsmit and George E. Uhlenbeck postulate that the electron has an intrinsic angular momentum, independent of its orbital characteristics

1927 W. Pauli adds the spin matrix term to the Schrödinger equation. The magnetic moment of the electron is paramagnetic.

1928 P.A.M. Dirac publishes his equation, of which that of Pauli-Schroedinger is a low energy limit.

1930 Schrödinger discovers Zitterbewegung for the electron in the Heisenberg representation of the Dirac equation.

1931 Dirac predicts the positron

1936 Euler Heisenberg NLED

1940 Pauli proves the spin-statistics theorem, discovered by Fermi in 1925

1948 Schwinger Anomalous Mag Moment;J Phys Rev 73, 416

1950 Tomonaga Bosonization of fermions in 1D Progr Theor Phys 5, 544

1951 Schwinger critical field Phys. Rev. 82, 664 (1951); 93, 615 (1954).

1959 Loudon, Am J Phys 27, 649; Elliott & Loudon, J. Phys. Chem. Solids 8, 382 Electrons in strong fields ($B > B_a = Z^2 m^2 e^3 c/\hbar^3 \simeq 2.35\ Z^2\ x\ 10^5\ [T]$) are 1D.

1963 Luttinger J. Math Phys 4, 1154

## Long solenoid, small quantum flux n

The anomalous magnetic moment can be neglected and

$$x^2 \cong \frac{2n}{2n+1} \quad ; \quad n \cong \frac{x^2}{1-x^2} \quad .$$

| n  | $x^2$   | N [m⁻³]       | $N_d = N\pi\lambda_c^2$ [m⁻¹] | $\lambda_c = \hbar/M_o\text{-}c$ [m] | d (N'=10¹¹) |
|----|---------|---------------|-------------------------------|--------------------------------------|-------------|
| 1  | 2/3     | 2.92•10³⁸     | 1.028•10¹⁴                    | 335•10⁻¹⁵                            | 972μ        |
| 10 | 20/21   | 1.57•10³⁸     | 3.91•10¹⁴                     | 889•10⁻¹⁵                            | 256 μ       |
| 30 | 60/61   | 9.54•10³⁷     | 6.94•10¹⁴                     | 1521•10⁻¹⁵                           | 144 μ       |
| 50 | 100/101 | 7.46•10³⁷     | 9•10¹⁴                        | 1963•10⁻¹⁵                           | 111 μ       |

Rojas had set $U=U_e=0$, while renormalizing the magnetic energy inside the states by the Euler-Heisenberg procedure. But $U_e = 0$ requires a very specific distribution of screening nuclei within and round the state. Let us evaluate the electrostatic as well as the unrenormalized magnetic energy per pair from a homogeneous charge distribution of screening nuclei *inside* the condensate, $U = E_m + U_e$. To this classical quantity, one may substract the renormalizing magnetic term. The total classical energy for this configuration has a local minimum at $n=11$. Without renormalization, none of the states below satisfy the condition for a charged condensate, i.e. $0 \leq -U < M_o\text{-}c^2$, especially since in the presence of screening nuclei $U_e$ is a minimum. The maximum electrostatic energy $U_{e\ max}$ would be that of an unscreened condensate in vacuum, which is clearly unphysical.

| n  | $x^2$   | $M_o$-    | $E_m$     | $-U_e$    | $M_o$-+ U |
|----|---------|-----------|-----------|-----------|-----------|
| 1  | 2/3     | 589.26keV | 294.63keV | 240.54keV | 643.35keV |
| 10 | 20/21   | 222keV    | 1.114MeV  | 915keV    | 421keV    |
| 30 | 60/61   | 129.7keV  | 1.96MeV   | 1.62MeV   | 468.7keV  |
| 50 | 100/101 | 100.5keV  | 2.54MeV   | 2.11MeV   | 528.7keV  |

## Long solenoid, large n

$$n = \frac{1}{2} \frac{x^2}{\left(\sqrt{1-x^2} - \gamma_A x^2\right)^2} .$$

| n | $(1-x^2)^{-1}$ | N [m$^{-3}$] | $N_d = N\pi\lambdabar_c^2$ [m$^{-1}$] | $\lambdabar_c = \hbar/M_o\text{-}c$ [m] | d (N'=10$^{11}$) |
|---|---|---|---|---|---|
| 100 | 194.62 | 5.4•10$^{37}$ | 1.27•10$^{15}$ | 2737•10$^{-15}$ | 78.6μ |
| 500 | 931.54 | 2.48•10$^{37}$ | 2.9•10$^{15}$ | 6109•10$^{-15}$ | 34.4 μ |
| 10$^3$ | 1808.7 | 1.78•10$^{37}$ | 4.17•10$^{15}$ | 8637•10$^{-15}$ | 24 μ |
| 10$^4$ | 14762.5 | 6.23•10$^{36}$ | 1.46•10$^{16}$ | 0.27Å | 6.85 μ |
| 10$^5$ | 86724.5 | 2.57•10$^{36}$ | 6.025•10$^{16}$ | 0.86Å | 1.66 μ |
| 10$^6$ | 286961 | 1.41•10$^{36}$ | 3.31•10$^{17}$ | 2.73Å | 301nm |
| 10$^7$ | 522625 | 1.05•10$^{36}$ | 2.45•10$^{18}$ | 8.63Å | 40.8nm |

If both the electrostatic and magnetic energies are counted for the large *n* long solenoids screened by a uniform inner charge distribution, the total energy peaks, then cancels and becomes negative after *n*=4495. This is clearly dependent upon the screening charge density: the mass of a negative energy condensate switches to positive if enough ions are expelled, and back to negative once new screening ions have settled inside. Without renormalization, states beyond the zero energy state, with *n* > 4495, can have a classical energy $U = E_m + U_e$ that satisfies the condition for charged states, given the proper distribution of screening charges. With the renormalization of the magnetic energy, more nuclei need to lie outside the condensate for the condition on *U* to be satisfied for such states, while smaller *n* is allowed.

| n | $(1-x^2)^{-1}$ | $M_o$- | $E_m$ | $-U_e$ | $M_o$-+ U |
|---|---|---|---|---|---|
| 100 | 194.6 | 72keV | 3.54MeV | 2.97MeV | 642.3keV |
| 500 | 931.54 | 32.3keV | 7.78MeV | 6.8MeV | 1.022MeV |
| 1000 | 1808.7 | 22.8keV | 10.86MeV | 9.76MeV | 1.12MeV |
|  |  |  |  |  |  |
| 4495 | 7298.6 | 10.8keV | 21.8MeV | 21.8MeV | 0 |
| 4496 |  |  |  |  |  |
|  |  |  |  |  | <0 |

The collapse continuing, we next have the

## Short solenoid, large n

$$n = \frac{2e}{hc^2} \Psi I$$

with $I$ as in (27) and

$$\Psi = 4\pi k N_d S = L/N' \quad , \quad S = \pi \lambdabar_c^2 \quad .$$

| $d/\lambdabar_c$ | 20 | 10 | 2 | 1 | 0.2 |
|---|---|---|---|---|---|
| k | ~1 | 0.9 | 0.6 | 0.5 | 0.2 |

| n | $(1-x^2)^{-1}$ | $N_d = N\pi\lambdabar_c^2$ [m$^{-1}$] | $\sqrt{N_S} = \sqrt{(N\lambdabar_c/2)}$ [m$^{-1}$] | $\lambdabar_c = \hbar/M_o\text{-}c$ [m] | d (N'=10$^{11}$) |
|---|---|---|---|---|---|
| 2.6•10$^7$ | 600000 | 6.82•10$^{18}$ | 2.7•10$^{13}$ | 1.47nm | 15nm |
| 4.6•10$^7$ | 650000 | 1.7•10$^{19}$ | 3.52•10$^{13}$ | 2.39nm | 5.91nm |
| 2.4•10$^8$ | 700000 | 8.4•10$^{19}$ | 4.95•10$^{13}$ | 5.42nm | 1.2nm |

## The condensate surface sheet has folded onto itself into an extreme ring

Now, $d=2\pi r$. For $N'=10^{11}$

| $(1-x^2)^{-1}$ | $N_{V'} = 4\pi N_S N_d/N'$ [m$^{-3}$] (N'=10$^{11}$) | $N_d = N\pi\lambdabar_c^2$ [m$^{-1}$] | $\sqrt{N_S} = \sqrt{(N\lambdabar_c/2)}$ [m$^{-1}$] | $N_p = N_S d$ [m$^{-1}$] (N'=10$^{11}$) | $\lambdabar_c = \hbar/M_o\text{-}c$ | d (N'=10$^{11}$) |
|---|---|---|---|---|---|---|
| 740000 | 4.9•10$^{40}$ | 1.3•10$^{22}$ | 1.73•10$^{14}$ | 2.32•10$^{17}$ | 68.5nm | 7.7pm |
| 741000 | 1.3•10$^{41}$ | 2.49•10$^{22}$ | 2•10$^{14}$ | 1.67•10$^{17}$ | 94nm | 4pm |
| 742000 | 5.6•10$^{41}$ | 6.57•10$^{22}$ | 2.6•10$^{14}$ | 1.03•10$^{17}$ | 154nm | 1.52pm |
| 743000 | 1.05•10$^{43}$ | 4.6•10$^{23}$ | 4.24•10$^{14}$ | 3.9•10$^{16}$ | 410nm | 2.14•10$^{-13}$m |
| 743500 | 2.15•10$^{45}$ | 1.61•10$^{25}$ | 1.03•10$^{15}$ | 6.57•10$^{15}$ | 2.41μ | 6.18•10$^{-15}$m |
| 743602 |  | 1.69•10$^{30}$ | 1.85•10$^{16}$ | 2.03•10$^{13}$ | 783μ | 5.9•10$^{-20}$m |

$N_{V'}$, the bulk density inside the folded sheet of the extreme ring is inversely proportional to $N'$ and large enough to allow nuclear reactions between heavy nuclei. With proper toroidal magnetic induction, the nuclear soup it encloses deconfines into nucleon or even quark condensates inside a collective bag that allows vanishing quark momenta. $N_{V'}$ ceases to be significant when the little radius shrinks to a few Fermi in favour of $N_p$, the perimetric density.

$n = 2eN'LI/hc^2$ with $I$ as in (27). For $N'=10^{11}$,

| $(1-x^2)^{-1}$ | $n$ | $M_o$- | $E_m$ | $-U_e$ | $U_{e\,max}$ |
|---|---|---|---|---|---|
| 740000 | $2.63 \cdot 10^7$ | $2.88eV$ | $86.67keV$ | $3.41GeV$ | $15GeV$ |
| 741000 | $1.95 \cdot 10^7$ | $2.1eV$ | $35.5keV$ | $2.46GeV$ | $11.75GeV$ |
| 742000 | $1.34 \cdot 10^7$ | $1.28eV$ | $9.3keV$ | $1.5GeV$ | $8.1GeV$ |
| 743000 | $6.16 \cdot 10^6$ | $480meV$ | $600eV$ | $569MeV$ | $3.7GeV$ |
| 743500 | $1.38 \cdot 10^6$ | $81meV$ | $3.9eV$ | $97MeV$ | $829MeV$ |
| 743602.3161… | 1 | $\sim 0.1\mu eV$ | $5 \cdot 10^{-18} eV$ | | $72.9eV$ |

$U_{e\,max}$ is the electrostatic energy for the unphysical unscreened state in vacuum and $U_e$ that for the similarly unphysical state completely screened by a uniform distribution of nuclei within. The classical unrenormalized energy can thus be cancelled or made suitably small, as might some smaller renormalized value, by choosing a distribution for the screening nuclei with energy between that for $U_e$ and $0$. Note the smallness of the classical magnetic energy $E_m$, as well as of $M_o$- for extreme $0^+$ states. The flux number $n$ decreases as $M_o$- $\to 0^+$ to its lower limit $n=1$. We thus have a giant electron-like object whose Compton radius spans several metres. The latter being proportional to $N'$, we recover the microscopic Compton radius for a single boson. In a dense magnetized lepton condensate, the neutrino acquires an effective mass and a tiny magnetic moment which, as computed by Rojas et al. are proportional to the lepton density, $\mu_\nu \sim 10^{-13} \mu_B$, $m_\nu \sim 250eV$ at $B_{c\,W}$- and $N \sim 10^{45} m^{-3}$. It follows that muon and $W$ extreme rings can be stable against decay. High magnetic fields could thus be substituted for high energy collisions to probe the Standard Model at very low energies.

## *The quantum degeneracy and fusion of nuclei*

The parameter of quantum degeneracy $A$ is

$$A_D = \frac{\hbar^D N_{nD}}{(2\pi m k_B T)^{D/2}}$$

where $D$ is the dimensionality and $N_{nD} = 2N_D/Z$ is the density of screening nuclei. A gas is said to be quantum degenerate when $A_D \geq 1$. For $A_D = 1$ and bosons of mass $m$, we have

$$T_D = \frac{\hbar^2 N_{nD}^{2/D}}{2\pi m k_B} \quad , \quad u_D = \frac{\hbar N_{nD}^{1/D}}{m\sqrt{\pi}}$$

For Fermions, the Fermi velocity $u$ would be

$$u_V = \frac{\hbar}{m}\left(\frac{3N_{nV}}{4\pi}\right)^{1/3} \quad ; \quad u_S = \frac{\hbar}{m}\left(\frac{N_{nS}}{\pi}\right)^{1/2} \quad ; \quad u_d = \frac{\hbar N_{nd}}{m}$$

where a factor $(2s+1)^{-1/D}$ was omitted owing to total polarization in the strong magnetic field. Since this is a maximal velocity, at constant density, a gas of degenerate bosons can have a higher temperature and a higher mean velocity. This increases the rate of fusion for bosons per unit volume (in the dimensionality $D$ considered) per second.

$$f = N_{nA} N_{nB} \langle \sigma_{AB} u \rangle$$
$$= \frac{1}{4} N_{nA}^2 \langle \sigma u \rangle_{A=B}$$

For the long solenoids, $N_V$ shows this rate to be huge, especially for light elements. Here, the fusion results only from the nuclear density. Its products, having energies generally beyond that of the confined Landau orbits of small $n$ solenoids, are ejected, surrounding them with a halo, while the confined nuclear gas is heated for large $n$ entities, except at the endings.

Assuming long solenoids to have 3D interiors and to be neutralized by identical nuclei with A, Z; we have, for small flux number $n=1,2,...$

$$T_V \cong 1.4 \cdot 10^7 \frac{2n}{(2n+1)^{3/2} A Z^{2/3}} [°K]$$

This decreases as $n$ increases and the state collapses, but can be reasonably high for small $n$ and light elements. Fermionic nuclei with a magnetic moment are less likely to condense as they will all be oriented the same way, necessitating an additional pairing mechanism in order to form vectors.

Similarly, the endings of long solenoids, the surface of short ones or of the first extreme tori are *2D*, with $N_S$ ranging from $10^{26}$ to $10^{29}$

$$T_s \cong 1.53 \cdot 10^{-19} \frac{N_s}{AZ} [°K]$$

Eventually, the extreme rings reach small radii below $10^{-14}m$, at which point the nuclear condensate is clearly one-dimensional, its lineal density $N_p$ being proportional to $N'$, the number of pairs and to $a$, comprised between $10^5$ and $10^2$. The corresponding temperature decreases as the critical field $B_c$ is reached

$$T_p \cong 3 \cdot 10^{-19} \frac{N_p^2}{AZ^2} \cong 3 \cdot 10^{-19} a^2 \frac{(N')^2}{AZ^2} [°K]$$

### *Magnetic binding of spinning nuclei or nucleons*

Typically, this is from tens to hundreds of eV. For a neutron, it is

$$532.4 x^2 [eV]$$

while the transition to a spin down state is twice that. The consequence is NMR spectroscopy in the soft x-ray range. NMR transitions shift a bound nucleon from one side to another of the current sheet of a short solenoid.

Another consequence is that spinning nuclei with opposite spins will seat on different sides of the sheet current of short solenoids while their charge keeps the nuclei on opposite sides apart; they will take on random orientations round extreme rings, breaking the degeneracy and the possibility of condensation, while scalar nuclei can sit on the sheet current of short solenoids or within the path of the extreme ring.

### *Landau confinement and transitions for nuclei*

The confinement of Landau orbits can occur when

$$r_L = \sqrt{\frac{2k\hbar}{ZeB}} < \frac{\hbar}{M_{o^-} c} = \lambda_C$$

$$r_L \cong 3.86 \frac{1}{x}\sqrt{\frac{k}{Z}} 10^{-13} [m]$$

which is true for *k=1* and Helium or higher Z nuclei and *B = 2B_c/3*. Any such confinement ceases at the endings. There will be Landau transitions between orbits at

$$E_k = \frac{Zek\hbar}{mc} \cong 557.5 \frac{Zkx^2}{A} [eV]$$

Since the Landau energy is inversely proportional to the mass, electrons in the *k>0* orbits within large *n* flux solenoids would have relativistic energies. The energy transfer from nuclear reactions is therefore improbable, as pointed out by Eric Lerner, and only the confined nuclear gas is heated, leaking from the endings.

## *An estimate for the initial energy of ions*

When the condensate forms as a *n=1* state, surrounding ions rush towards it to screen it. Taking a long *n=1* solenoid, its charge per meter will be $2eN_d$ with $2N_d/Z$ screening nuclei. Assuming that these come from a cylinder of radius

$$R \cong \sqrt{\frac{2N_d A}{\pi \rho n_A Z}}$$

Using the values for *R*, the compression ratio *k* is

$$k = \frac{R^2}{\lambda_c^2} \quad , \quad T_f = T_o k^{\gamma-1} \quad .$$

Taking the adiabatic index $\gamma = 5/3$ and $\gamma = 4/3$,

|  | $T_o$ | R | k | $T_{f\gamma=5/3}$ | $U_{\gamma=5/3}$ | $T_{f\gamma=4/3}$ | $U_{\gamma=4/3}$ |
|---|---|---|---|---|---|---|---|
| liquid $^1H_1$ | 20°K | 39.4nm | $1.385 \cdot 10^{10}$ | 115M°K | 9.935keV | 48k°K | 4eV |
| solid $^{26}Fe_{56}$ | 300°K | 4.96nm | $2.198 \cdot 10^8$ | 109M°K | 9.4keV | 181k°K | 15.6eV |

The dynamics of condensation, final energies and temperatures depend upon $\gamma$. The value $\gamma = 5/3$ governs many non-relativistic systems: noble gases, the sun, tokamaks, inertial confinement fusion, etc. The final temperature of the nuclei will be lower if atoms or groups of them condense separately in the magnetized region, the condensate bits then joining by monopolium annihilation, as well as when some energy from the nuclei is transmitted to non-condensed electrons, especially when the initial energy is too high for confinement and the nuclei oscillate in and out of the condensate until they settle within it. The very existence of electron condensates suggests smaller nuclear temperatures and $\gamma < 5/3$.

## *The evolution of electronic condensates*

Although the condensate may be charged, the binding of its bosons must prevail. Therefore, the condensate needs to be stabilized at the outset by an abundant supply of cold screening nuclei with the proper reactivity so that $-U \in \left(0, M_{o^-} c^2\right[$. A pulsed electric arc in a cool, flowing, dense liquid medium, so that the condensates are diluted within it, in a vessel that allows the acoustic overpressure following cavitation to be felt as the condensate forms and evolves is therefore desirable. While the gas of nuclei within a long solenoid can be considered to be three-dimensional, its density in a fully screened condensate is so high that, for light elements whose magnetic polarization is suitable, fusion is quasi instantaneous. If it were to proceed at too high or too low a rate, the proper screening would be destroyed, as would the state. The limiting factor, both for the stabilization of the condensate by screening and for the fusion of light elements to proceed is the rate at which cool ions are supplied and at which they can settle within the condensate, versus that at which they are ejected from it by nuclear reactions. In the initial stages of loading, the nuclei form a dilute screened 3D gas inside the filament. Boson nuclei are preferable because of their lower pressure than Fermions, which therefore allows for faster loading. At some point, the density of bosons ceases to screen them from one another, their mutual interactions predominate, the gas of nuclei is fermionized, its pressure increases, the loading stops while the fusion rate soars. Initially, the small $n$ long solenoids have no confining Landau orbits for the products of fusion so that the latter leak out: the newly formed filament has a uniformly hot outer halo of ions while stabilization and cooling are critical. At this stage, the fusion is mostly density-driven. The "poisoning" of nuclear transformations by pure deuterated water could be due to the too great reactivity of deuterium, resulting in the destruction of the condensate from the ejection of fusion products and subsequent underscreening. For $n$ between $10^4$ and $10^7$, the long solenoid acquires confining Landau orbits for the charged products of fusion and operates as a pulsed thermonuclear reactor whose cycle comprises the loading of a Bose gas from its dense surroundings, its fermionization, fusion, thermonuclear heating, and the emission of the hot byproducts of fusion from the open endings as puffs of energetic ions. This would be the explanation for the caterpillar-like traces of "strange radiation" seen by L. Urutskoev et al. Suitable light ions for fusion in a high magnetic field in the bulk of long solenoids would be Deuterium in combination with either $He_4$, $Li_6$ or $C_{12}$, the last two being preferable because they are liquids above room temperature. In that case, the concentration of the reacting fuel must be suitably low to preserve the screening. As the solenoid collapses to the short form, the system becomes two-dimensional and two-sided, *i.e.* there is a charged current sheet with opposing magnetic field orientations on each side. This facilitates the fusion of nuclei with opposite polarizations, such as *D-D*. Finally, the screening allowed the previous states to survive, there is the extreme ring. Initially, the charge density within the folded sheet is so high as to confine even heavy elements into a nuclear soup. Because of their reactivity, the light atoms are likely to have fused or have been ejected. The transformation seems to favour the most common even-even nuclei, so that bosonization is possible both for protons and neutrons. For elements close to Iron, where the binding energy is at a maximum, the exchange interactions result in transformations in which this energy must remain constant, as observed by Urutskoev et al, while there is no allowed transformation for Iron and Nickel themselves, since these maximize the binding energy in the periodic table. As the peripheral charge density decreases, the soup condenses again into the transformed nuclei. In the narrowed condensate, boson nuclei can only be fermionized. At the very end, because of the milder peripheral charge density, any screening nuclei are both fermionized and separated enough from one another to form inert pseudo-molecular magnetized ring chains. These have both a magnetic moment and angular momentum; they precess in an external magnetic field in a low amplitude mechanical wobble. The rate for an extreme ring screened by *(Z,A)* nuclei with *B* in Gauss is

$$\nu = \frac{g_e}{h}\frac{Ze\hbar B}{2m_n c} \cong \frac{ZB}{A}1527\,[Hz]$$

I suspect that the correlated traces observed by Ivoilov are cross-sections of just such a precessing ring, with some drift caused by residual charge. As soon as $M_o\text{-}\leq\gamma_A M$, a transition may occur to a negative mechanical mass, which should appear as the dynamics of an opposite charge.